# Mid-infrared continuous wave parametric amplification in tapered chalcogenide microstructured fibers

Sida Xing[1], Davide Grassani[1], Svyatoslav Kharitonov[1], Laurent Brilland[2], Céline Caillaud[2], Johann Trolès[3], Camille-Sophie Brès[1,*]

[1]Ecole Polytechnique Fédérale de Lausanne, Photonic Systems Laboratory (PHOSL), STI-IEL, Station 11, CH-1015 Lausanne, Switzerland;

[2]SelenOptics, 57 rue de La Girolle; 35170 Bruz, France

[3]Equipe Verres et Céramiques, UMR-CNRS 6226, Institut des Sciences Chimiques de Rennes, Université de Rennes1, 35042 Rennes Cedex, France

*Corresponding author: Phone +41 21 693 7866; Fax +41 21 693 6877;

Email: camille.bres@epfl.ch

**Abstract:** As photon mixing is not inherently limited to any specific spectral region, parametric processes represent a compelling solution for all-optical signal processing in spectral windows not easily accessible by other technologies. Particularly, the continuous-wave pumping scheme is essential for any application requiring modulated signals or precise spectroscopic characterization. Highly nonlinear fibers enabled record performances for wavelength conversion and amplification in the telecommunication band, however no waveguiding platforms have yet solved the trade-off between high-nonlinearity, low propagation losses and dispersion in the mid-infrared. Here, we show mid-infrared continuous-wave parametric amplification in a $Ge_{10}As_{22}Se_{68}$ fiber. Leveraging state-of-the-art fabrication techniques, a novel tapered photonic crystal fiber geometry enabling 4.5 dB signal amplification and 2 dB idler conversion efficiency is experimentally demonstrated using only 125 mW of pump in the 2 μm wavelength range. This result is not only the first ever continuous-wave parametric amplification measured at 2 μm, in any waveguide, but also establishes GeAsSe PCF tapers as the most promising all-fibered, high efficiency continuous-wave parametric converter for advanced applications in the mid-infrared.

**Introduction**

Parametric wavelength conversion and amplification based on four-wave mixing (FWM) pumped in continuous wave (c.w.) regime allows for the preservation of narrow linewidth and any complex modulation. Compared with other mid-infrared (MIR) light generation/amplification techniques, parametric conversion in waveguides provides a compact, widely tunable, highly efficient solution operating at room temperature[1, 2, 3]. The complete transparency to any bit rate, together with reduced self- and cross-phase modulation experienced by the c.w. pump, enables record performances of all-optical processes in terms of speed, bandwidth and noise reduction in the telecommunication band by exploiting highly nonlinear silica fibers[4]. Recently, the availability of tunable, c.w. watt-level fiber lasers has been extended from the telecommunication band ($\lambda$ = 1550 nm) to the onset of the MIR ($\lambda \approx$ 2000 nm) by using thulium (Tm) and holmium (Ho) doped silica fibers as active media[5, 6]. Such lasers can potentially be used for pumping nonlinear waveguides, translating the benefits of all-guided c.w. parametric conversion and amplification to MIR applications like precise spectroscopy[7], free-space communication[8], and remote sensing[9]. However, adequate platforms are still yet to be shown.

Waveguide engineering for c.w. parametric amplification should take material, geometry and fabrication, into consideration due to the sensitivity of the process on nonlinear strength, phase matching and losses. To reference these requirements, we use a figure of merit (FOM)[10] defined as the ratio between nonlinear coefficient $\gamma$ and the propagation loss $\alpha$. Ideally, this FOM should be as large as possible. While silica highly nonlinear fibers exhibit too high absorption from 2 micron and a too low $\gamma$ to have good performances in MIR, materials possessing both high nonlinear refractive indices ($n_2$) and good MIR transparency include, in particular, soft glasses (tellurite and chalcogenide glasses) and silicon. Up to now, platforms with highest $\gamma$, such as silicon waveguides and $As_2Se_3$ tapers, exhibit high propagation losses, in the order of dB/cm. They are either due to sidewall scattering and multi-photon absorption[1, 11], or absorption in polymer coating[12], significantly impacting their FOM. As such, only pulsed pumping schemes were demonstrated. Soft glasses photonic suspended core fibers (SCFs) do not need any polymer coating and in recent years[13, 14, 15] they have been fabricated with low loss. Due to the large refractive index difference, higher order modes can be easily excited in SCFs, seriously affecting the phase matching condition[14]. The large air holes also allow extended contact area with atmosphere such that strong water absorption and crystallization were found to severely degrade their nonlinear optical performance[16]. On the other hand, PCF structure is more promising due to the possibility of single mode behavior over large wavelength span and better immunity to the environment. We have tested a GeAsSe PCF[17], which showed low loss, single mode operation and high FOM under CW pumping at telecom and 2um range, leading to a relatively large FOM. The strong normal material dispersion of chalcogenides represents however a major hurdle often translating into a tradeoff between dispersion, losses and multimode features.

|  | Silica HNLF[18, 19] | Si chip[1, 11] | $As_2S_3$ chip[20] | $As_2Se_3$ taper[12, 21] | Tellurite SCF[13] | GeAsSe PCF[17] | GeAsSe PCF tapered |
|---|---|---|---|---|---|---|---|
| $n_2$ | $2.7 \cdot 10^{-20}$ | $1.1 \cdot 10^{-17}$ | $3 \cdot 10^{-18}$ | $1.1 \cdot 10^{-17}$ | $5.9 \cdot 10^{-19}$ | $5.3 \cdot 10^{-18}$ | $5.3 \cdot 10^{-18}$ |
| $n_2/\alpha$ $m^2/(W \cdot dB)$ | $3.6 \cdot 10^{-19}$ | $4.3 \cdot 10^{-20}$ | $6 \cdot 10^{-20}$ | $3.7 \cdot 10^{-19}$ | $2 \cdot 10^{-18}$ | $5.3 \cdot 10^{-18}$ | $1 \cdot 10^{-17}$ |
| FOM $(dB \cdot W)^{-1}$ | $3.2 \cdot 10^{-2}$ | 0.44 | 0.198 | 0.8 | 0.56 | 1.7 | **19.78** |

**Table 1 | Figure of merit (FOM) of various nonlinear platforms used for parametric processes at 2μm.** Upper limits are calculated based on the information provided by the authors in the referenced publications. The work presented here is highlighted in the last column, resulting in a figure of merit a couple orders of magnitude higher than other platforms.

Here we overcome this tradeoff by dispersion engineering a GeAsSe fiber through the combination of microstructuring and tapering. The fiber is fabricated by the molding technique, allowing large effective length and low losses. Tapering also leads to an increased $\gamma$, resulting in a FOM several orders of magnitude higher than other approaches, as summarized in table 1. With this tapered PCF, we show for the first time parametric amplification and conversion efficiency (CE) above the transparency threshold using a c.w. pump laser in MIR region.

## Material and Methods
### Fiber design and fabrication

The zero dispersion wavelength (ZDW) of a PCF is highly dependent on the diameter-to-pitch ratio $(\rho)$[22]. For a given core diameter $\phi$, a large $\rho$ value blue shifts the ZDW but also increases the multimode behavior and fabrication difficulty of the fiber. Within fabrication limits ($\rho \leq 0.6$) of GeAsSe PCF, solely varying $\rho$ does not allow for sufficient shift of the ZDW towards the 2 µm thulium/holmium band. Additional reduction of the core size is necessary to enhance the waveguide dispersion. Tapering is an efficient way to reduce the fiber waist size to values not easily accessible with direct fabrication (Fig. 1a). The designed fiber was thus fabricated in a two-step method. First, a GeAsSe PCF featuring three rings of holes aiming at $\rho \approx 0.6$ and a $\phi_s \approx 4$ µm diameter solid core was fabricated (Fig. 1b). This initial geometry was chosen for three reasons, mainly related to fabrication: a) the small initial core facilitates the tapering step, b) GeAsSe allows for the reduction of the core size without excess loss and c) it has a lower material dispersion with respect to AsSe compounds[23]. Second, the central part of the fiber was tapered to achieve a core size diameter of about 1.5 µm (Fig. 1c).

The fabrication starts from a previously synthetized highly purified $Ge_{10}As_{22}Se_{68}$ bulk glass. The glass has an optical loss of 0.6 dB/m measured at 1.55 µm. Part of the GeAsSe glass was then modeled to fabricate a preform with 3 hexagonal rings of air holes[24]. The preform was drawn into a cane of 4 mm diameter. This cane was then inserted into an $As_2Se_3$ tube and the whole tube assemble was drawn into a photonic crystal fiber with an outer diameter of approximately 130µm and a core size about 4µm. The pitch distance, Λ, was measured to be approximatively 2.88 µm and the air hole diameter was 1.69 µm; leading to a diameter-to-pitch ratio of 0.58. The propagation loss of the PCF was measured to be 0.65 dB/m at 1.55 µm from a cutback experiment. Using a segment of this PCF, tapered fibers in 1m length range were then fabricated on the drawing tower at a temperature around 270 °C. The core diameter in the waist portion was decreased to approximately 1.5 µm (see Fig. 1c).

**Experimental setup**

While an interferometric method was used to characterize the dispersion of the un-tapered PCF, because of the phase delay contribution from the un-tapered and transition sections, it cannot be directly applied to measure the dispersion of the tapered PCF as it is in Fig. 1a. The four-wave mixing (FWM) process, at basis of parametric amplification and frequency conversion experiments, can be used also to retrieve the dispersion, giving a good indication of fiber uniformity, suppression of the higher order modes and strength of birefringence.

We used an all-fibered FWM set-up, detailed in Fig 2. The 2 µm pump and signal lasers used in this experiment were implemented with Tm-doped fiber pumped by L-band erbium doped fiber amplifier (EDFA). The pump laser utilized either linear or ring cavity depending on the wavelength. To achieve signal amplification, a high slope efficiency, linear cavity laser was built with a 1950 nm fiber Bragg grating (FBG) pair. Due to a lack of matched FBG pairs at other wavelengths, a ring cavity was built and different FBGs were used to select the pump wavelength for dispersion characterization. Both pump configurations were checked using a photodetector to confirm no existence of pulsing operation. The pump laser linewidth was measured to be approximately 0.08 nm at 1950 nm in the linear cavity configuration. The signal laser was based on a ring cavity where the wavelength was selected by a tunable bandpass filter of 1 nm linewidth. The length of the Tm-doped fiber was 11.5 m in the pump and 4.5 m in the signal laser. The output from pump and signal lasers were combined by a 95/5 fiber coupler, designed for operations at 2 µm, and then send through a circulator to prevent back reflection. A fibered polarization beam splitter was mounted after the circulator. One arm of the polarizer was connected to a polarization maintaining (PM) lensed fiber for coupling into the PCF and the other arm was used for power monitoring. The PM lensed fiber was fabricated with PM1550 and has a beam spot diameter of approximately 4 µm at 2 µm wavelength. The output from the chalcogenide fiber was collected using another PM lensed fiber with the same parameters. The lensed fiber was then connected to an optical spectrum analyzer (OSA), Yokogawa AQ 6375, for data recording. A total insertion loss of 8.5 dB was measured at 1950 nm, where approximately 4.5 dB are the total coupling losses, 3.4 dB comes from the transition regions and 0.6 dB comes from propagation losses ($\alpha$).

**Result and discussion**

We experimentally measured the dispersion of the un-tapered, 4 µm core PCF using an interferometric technique[25]. Excellent agreement with simulation indicates the accuracy of our model and quality of the fiber. As expected the dispersion is strongly normal at 2 µm, with a ZDW located near 2.9 µm. Building form this result, we compute the dispersion of the tapered fiber as shown in Fig. 3a, with shaded area representing possible birefringence coming from an approximately 2% fabrication error in the holes diameter. The ZDW is expected between 2.1 and 2.2 µm, depending on the input laser polarization. While simulations indicate a slight multimode behavior, the transition region acts as a mode filter so that only the fundamental mode propagates in the taper waist region. Very low losses in the waist region were measured to be approximately 0.5 dB/m at 2 µm.

The parametric behavior of the tapered fiber was studied by FWM using a c.w. pump together with a c.w. signal. To reduce the impact of other nonlinear effects and avoid potential fiber damage, all characterizations were performed at low pump power level (< 20 mW in fiber waist). Idler waves showed well-defined phase matching features, confirming the uniformity, single mode nature and low loss of the taper. Nonlinear Schrodinger equation (NLSE) was then used to fit the CE data in order to retrieve the fiber parameters. Since the pump laser is narrow linewidth and low power, the simplified NLSE was used for dispersion fitting:

$$i\frac{\partial}{\partial z}A(z,t) + \frac{i\alpha}{2}A - \frac{\beta_2}{2}\frac{\partial^2 A}{\partial t^2} + \gamma |A|^2 A = 0 \ . \quad (1)$$

In the above equation, $A(z,t)$ is the electric field amplitude; $\beta_2$ is the group velocity dispersion (GVD); $\alpha$ is the linear propagation loss. The nonlinear coefficient $\gamma$ of the tapered PCF was retrieved to be $\gamma \approx 10 \ W^{-1}m^{-1}$, in agreement with the value calculated using the measured $n_2$ from our previous work[17] and the effective area $A_{eff}$ simulated with the finite element method (FEM) program COMSOL. Optimization was performed to fit the positions of the experimental CE dips/minima, which give also another estimate of the propagation loss. The tolerance of this fitting was set to 0.05 nm. The contribution of the un-tapered input side was also considered and included in the simulated result of the tapered part. As we will show, due to the larger effective area and the shorter length, the effect of the un-tapered region mainly resulted in a slightly rise of the magnitude of the CE minima, without changing their position. We extracted the dispersion parameter $\beta_2$ of the waist region for pump wavelengths at 1950 nm, 1980 nm, 2008 nm and 2040 nm, while the signal was tunable over 140 nm in the 2 μm range, and for different tapered fibers, with waist lengths ranging from 0.8 m to 1.3 m, fabricated using the same fiber spool. The experimental values for all fibers are in excellent agreement with the predicted dispersion (Fig. 3a). An example of FWM spectra for a 2040 nm pump with 13 mW of coupled pump power and approximately 1 mW of signal power is shown in Fig. 3b.

The fiber birefringence was experimental confirmed in our experiment. An ideal hexagonal PCF is not birefringent[26], however, the symmetry of real PCF is always broken due to the fabrication imperfections. The slight variations of the cladding air hole size lead to effectively a different diameter-to-pitch ratio depending on the input light polarization. From our numerical simulation shown in Fig. 3c, the discrepancy becomes more evident when going closer to the ZDW. Thus, we checked the fiber birefringence experimentally at 2040 nm by rotating the angle of input Polarization maintaining (PM) lensed fiber. The recorded four wave mixing (FWM) spectra were converted as CE versus signal detuning curve. We found the effective fast and slow axis of this fiber, shown in Fig. 3c, clearly showing the fiber is birefringent. This polarization dependent dispersion corresponds to fluctuation of $\rho$ of less than 1%, further proving the good uniformity of the tapered PCF.

The efficiency of the parametric process, as a result of the combination of high $\gamma$ and long effective length, leads to amplification of the laser ASE, as depicted in Fig. 4a for a 1950 nm pump. Phase matching features emerged from the OSA noise floor increasing the pump power. With only 125 mW pump coupled, signal amplification and cascaded FWM up to 3 orders were observed while maintaining the initial linewidth of the signal (Fig. 4b). Since our pump laser linewidth is much smaller than the signal laser, its contribution to the idler linewidth is negligible. The idler conversion efficiency (CE), defined as idler output power over input signal power, is plotted in Fig. 4c as a function of signal detuning when 125 mW were coupled in taper waist at 1950 nm. At the idler side, operation above transparency is observed near the pump with a -20 dB bandwidth of 15 nm, while we recorded up to 4.5 dB of signal amplification.

In Fig. 5a, a reduction of the amplification bandwidth can be seen with increasing pump power. The On/OFF amplification was calculated using the ration between signal power at input of the fiber waist and the signal power at the output of the fiber waist. The narrowing of amplification bandwidth is due to the nonlinear contribution to the phase mismatch as the pump is in the normal dispersion regime. To quantize the effect of nonlinear phase contribution and for comparison with the theory, we also plotted the theoretical position of the first dip as a function of the pump power as shown in Fig. 5b. A perfect matching with the theory indicated a good estimation of linear and nonlinear phase mismatch and confirmed no change in the fiber geometry even at the highest pump power.

The input-output characteristics of the tapered fiber pumped at 1950 nm were further examined with a signal at 1951.5 nm. The CE as a function of pump power is plotted in Fig. 6a. The fitted experimental data exhibits a slope of approximately 1.8 with no onset of saturation. Form coupled wave equation theory, in the undepleted pump approximation, we expect a quadratic relation between pump and idler. This small discrepancy in our experiment comes from the cascaded FWM when the idler acts as a pump and is depleted due to the high power of the pump which now acts as the signal. Fig. 6b reproduces the expected theoretical relation between CE and signal amplification ($A_s$) given by $A_s=1+CE$, excluding the involvement of luminescence or non-parametric processes. It is also worth noting that our results establish the excellent power handling of the chalcogenide fiber in MIR. Before 2015, the maximum sustained c.w. intensity reported was 125 kW/cm$^2$ at 5.6 µm (ref. 27). Recently, damage threshold intensity was found to be higher than 5.6 MW/cm$^2$ when tested with a thulium doped fiber laser at 1.97 µm (ref. 28). Here we confirmed these findings with an even higher laser intensity of at least 6.62 MW/cm$^2$. Also, shown in Fig. 6a is the relationship between input and output pump power, as monitored on the OSA: the linear fit indicates constant fiber losses and stable coupling. To further confirm that no irreversible structural damage was induced during the 125 mW coupled pump power test, the fiber was re-characterized right after with 3 mW of coupled pump. The parametric behavior of the fiber remained identical, while no structural or losses changes were observed as seen in Fig. 6c.

To check the repeatability of the results, the two additional tapered PCFs with nominal identical geometry were also characterized. The two additional tapered PCFs have lengths of 0.8 and 1.2 m. Since the length of the three tapers differed slightly, CE over length squared is plotted as a function of coupled pump power in Fig. 6d. CE transparency and signal amplification were observed in all tapers, even at different pump wavelengths, confirming the repeatability of the experiment and of the taper features. The slight discrepancies are due to small loss variations in the taper transition region. To reach higher amplification, the pump power was increased leading to a damage of the fiber propagating backward to the input end. The damage is not due to power handling limitations but rather to a fiber fuse from a localized defect, likely caused by microbending resulting in light leakage in the coating which strongly absorbs 2 µm light (see Supplementary material). It should be noted that fiber fuse has been observed in silica fibers for similar power densities (MW/cm$^2$ range).

**Conclusions**

In summary we presented for the first time parametric amplification at 2 µm using low c.w. pumping power and relatively short fiber. The engineered GeAsSe PCF taper combines low dispersion (ZDW near 2.1 µm), high $\gamma$ of 10 W$^{-1}$m$^{-1}$ and low losses of 0.5 dB/m. Up to 4.5 dB of amplification was measured with only 125 mW of power inside the waist region. The fiber showed excellent power handling capabilities answering persistent doubts over c.w. operation in chalcogenide fibers. Moreover, they can be spliced with silica fibers[28] enabling compatibility with current fiber based devices and reducing coupling losses. Longer fiber and better polymer protection layer process (see Supplementary I) will lead to better CE and signal amplification. The wavelength dependent loss of GeAsSe[29] shows absolutely no obstacles towards pumping at the ZDW or slightly in the anomalous regime, which would further increase efficiency and bandwidth. We believe that these results are a breakthrough, paving the way for fully fibered optical processing devices based on parametric processes for the MIR range operating on low c.w. power.

**Acknowledgements**

This work was supported by the European Research Council under grant agreement ERC-2012-StG 306630-MATISSE. The authors are thankful to Adrien Billat from Photonics Systems Laboratory, École Polytechnique Fédérale de Lausanne for taking the scanning electron microscope (SEM) images of the fused fiber.

**Author contributions**

S.X. and D.G performed the experiments. S.X, D.G and S.K. carried modeling and theoretical analysis. S.K. performed the dispersion measurements based on interferometric method. L.B., C.C. and J.T. fabricated the fiber. C.S.B. supervised experiments. All authors contributed to the writing of the manuscript.

**References**

1. Zlatanovic S, Park JS, Moro S, Boggio JMC, Divliansky IB, AlicNikola, *et al.* Mid-infrared wavelength conversion in silicon waveguides using ultracompact telecom-band-derived pump source. *Nat Photon* 2010, **4**(8)**:** 561-564.

2. Lau RKW, Ménard M, Okawachi Y, Foster MA, Turner-Foster AC, Salem R, *et al.* Continuous-wave mid-infrared frequency conversion in silicon nanowaveguides. *Optics Letters* 2011, **36**(7)**:** 1263-1265.

3. Liu X, Kuyken B, Roelkens G, Baets R, Osgood RM, Green WMJ. Bridging the mid-infrared-to-telecom gap with silicon nanophotonic spectral translation. *Nat Photon* 2012, **6**(10)**:** 667-671.

4. Hansryd J, Andrekson PA, Westlund M, Li J, Hedekvist P-O. Fiber-based optical parametric amplifiers and their applications. *IEEE Journal of Selected Topics in Quantum Electronics* 2002, **8**(3)**:** 506-520.

5. Simakov N, Hemming A, Clarkson WA, Haub J, Carter A. A cladding-pumped, tunable holmium doped fiber laser. *Optics Express* 2013, **21**(23)**:** 28415-28422.

6. Li J, Sun Z, Luo H, Yan Z, Zhou K, Liu Y, *et al.* Wide wavelength selectable all-fiber thulium doped fiber laser between 1925 nm and 2200 nm. *Optics Express* 2014, **22**(5)**:** 5387-5399.

7. Wysocki G, Lewicki R, Curl RF, Tittel FK, Diehl L, Capasso F, *et al.* Widely tunable mode-hop free external cavity quantum cascade lasers for high resolution spectroscopy and chemical sensing. *Appl Phys B* 2008, **92**(3)**:** 305-311.

8. Luzhansky E, Choa F-S, Merritt S, Yu A, Krainak M. Mid-IR free-space optical communication with quantum cascade lasers. 2015; 2015. p. 946512-946512-946517.

9. Kameyama S, Imaki M, Hirano Y, Ueno S, Kawakami S, Sakaizawa D, *et al.* Development of 1.6 μm continuous-wave modulation hard-target differential absorption lidar system for CO2 sensing. *Optics Letters* 2009, **34**(10)**:** 1513-1515.

10. Marhic ME, Andrekson PA, Petropoulos P, Radic S, Peucheret C, Jazayerifar M. Fiber optical parametric amplifiers in optical communication systems. *Laser & Photonics Reviews* 2015, **9**(1)**:** 50-74.

11. Liu X, Osgood RM, Vlasov YA, GreenWilliam MJ. Mid-infrared optical parametric amplifier using silicon nanophotonic waveguides. *Nat Photon* 2010, **4**(8)**:** 557-560.

12. Abdukerim N, Li L, Rochette M. Chalcogenide-based optical parametric oscillator at 2 μm. *Optics Letters* 2016, **41**(18)**:** 4364-4367.

13. Cheng T, Zhang L, Xue X, Deng D, Suzuki T, Ohishi Y. Broadband cascaded four-wave mixing and supercontinuum generation in a tellurite microstructured optical fiber pumped at 2 μm. *Optics Express* 2015, **23**(4)**:** 4125-4134.

14. Le SD, Nguyen DM, Thual M, Bramerie L, Costa e Silva M, Lenglé K*, et al.* Efficient four-wave mixing in an ultra-highly nonlinear suspended-core chalcogenide As38Se62 fiber. *Optics Express* 2011, **19**(26)**:** B653-B660.

15. Savelii I, Mouawad O, Fatome J, Kibler B, Désévédavy F, Gadret G*, et al.* Mid-infrared 2000-nm bandwidth supercontinuum generation in suspended-core microstructured Sulfide and Tellurite optical fibers. *Optics Express* 2012, **20**(24)**:** 27083-27093.

16. Mouawad O, Amrani F, Kibler B, Picot-Clémente J, Strutynski C, Fatome J*, et al.* Impact of optical and structural aging in As2S3 microstructured optical fibers on mid-infrared supercontinuum generation. *Optics Express* 2014, **22**(20)**:** 23912-23919.

17. Xing S, Grassani D, Kharitonov S, Billat A, Brès C-S. Characterization and modeling of microstructured chalcogenide fibers for efficient mid-infrared wavelength conversion. *Optics Express* 2016, **24**(9)**:** 9741-9750.

18. Milam D, Weber MJ. Measurement of nonlinear refractive-index coefficients using time-resolved interferometry: Application to optical materials for high-power neodymium lasers. *Journal of Applied Physics* 1976, **47**(6)**:** 2497-2501.

19. Gershikov A, Shumakher E, Willinger A, Eisenstein G. Fiber parametric oscillator for the 2 μm wavelength range based on narrowband optical parametric amplification. *Optics Letters* 2010, **35**(19)**:** 3198-3200.

20. Lamont MRE, Luther-Davies B, Choi D-Y, Madden S, Gai X, Eggleton BJ. Net-gain from a parametric amplifier on a chalcogenide optical chip. *Optics Express* 2008, **16**(25)**:** 20374-20381.

21. Lenz G, Zimmermann J, Katsufuji T, Lines ME, Hwang HY, Spälter S*, et al.* Large Kerr effect in bulk Se-based chalcogenide glasses. *Optics Letters* 2000, **25**(4)**:** 254-256.

22. Russell PSJ. Photonic-crystal fibers. *Journal of Lightwave Technology* 2006, **24**(12)**:** 4729-4749.

23. Toupin P, Brilland L, Trolès J, Adam J-L. Small core Ge-As-Se microstructured optical fiber with single-mode propagation and low optical losses. *Opt Mater Express* 2012, **2**(10)**:** 1359-1366.

24. Coulombier Q, Brilland L, Houizot P, Chartier T, N'Guyen TN, Smektala F*, et al.* Casting method for producing low-loss chalcogenide microstructured optical fibers. *Optics Express* 2010, **18**(9)**:** 9107-9112.

25. Kharitonov S, Billat A, Brès C-S. Kerr nonlinearity and dispersion characterization of core-pumped thulium-doped fiber at 2 μm. *Optics Letters* 2016, **41**(14)**:** 3173-3176.

26. Steel MJ, White TP, Martijn de Sterke C, McPhedran RC, Botten LC. Symmetry and degeneracy in microstructured optical fibers. *Optics Letters* 2001, **26**(8)**:** 488-490.

27. Sanghera JS, Shaw LB, Aggarwal ID. Chalcogenide Glass-Fiber-Based Mid-IR Sources and Applications. *IEEE Journal of Selected Topics in Quantum Electronics* 2009, **15**(1)**:** 114-119.

28. Thapa R, Gattass RR, Nguyen V, Chin G, Gibson D, Kim W*, et al.* Low-loss, robust fusion splicing of silica to chalcogenide fiber for integrated mid-infrared laser technology development. *Optics Letters* 2015, **40**(21)**:** 5074-5077.

29. Toupin P, Brilland L, Méchin D, Adam J-L, Troles J. Optical Aging of Chalcogenide Microstructured Optical Fibers. *Journal of Lightwave Technology* 2014, **32**(13)**:** 2428-2432.

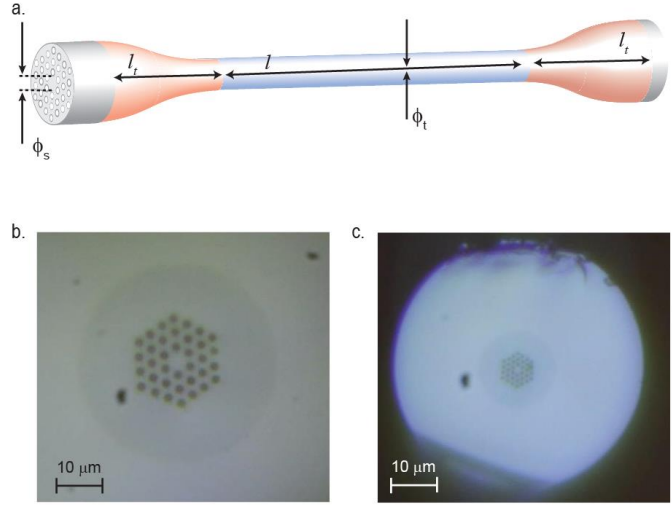

**Figure 1 | Dispersion engineered GeAsSe tapered PCF. a.** schematic of the GeAsSe fiber: the microstructured fiber of initial core size $\phi_s = 4$ μm is tapered to a new core size $\phi_t = 1.5$ μm over a length $l$. On either side of the taper is a $l_t = 1.5$ cm waist region acting as a mode filter. **b.** Image of the fabricated microstructure fiber before tapering. **c**. Image of the fiber after tapering, within the waist. The structure appears well preserved.

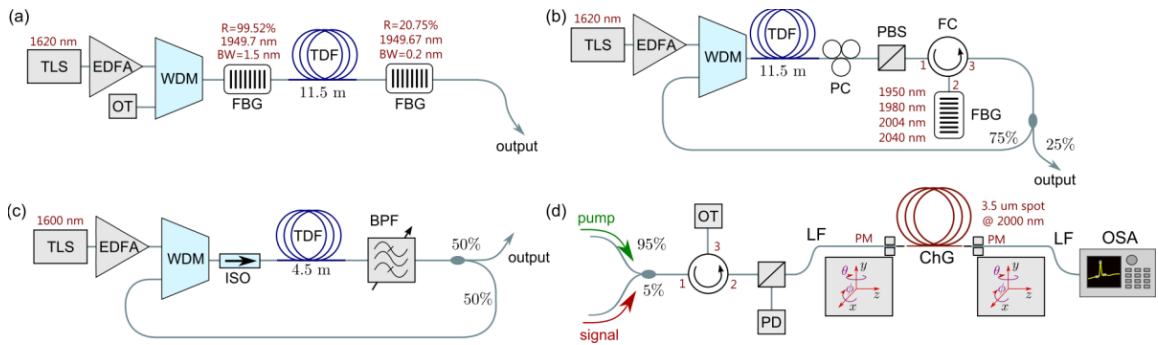

**Figure 2 | Experimental setup for GeAsSe PCF characterization.** TLS: tunable laser source; WDM: wavelength division multiplexer; TDF: Tm-doped fiber; OT: optical terminator; ISO: isolator; PC: polarization controller; BPF: bandpass filter; PBS: polarization beam splitter; PD: photodetector (2μm); LF: lens fiber; **a.** Pump laser cavity used to get parametric amplification in our experiment, which favors high slope efficiency and simple geometry to avoid pulsing; **b.** Pump laser cavity for low power characterization and dispersion retrieval of ChG fiber. An additional PC was inserted at the output of coupler before going to the PBS. **c.** Tunable signal laser; **d.** Complete setup for light injection and data recording.

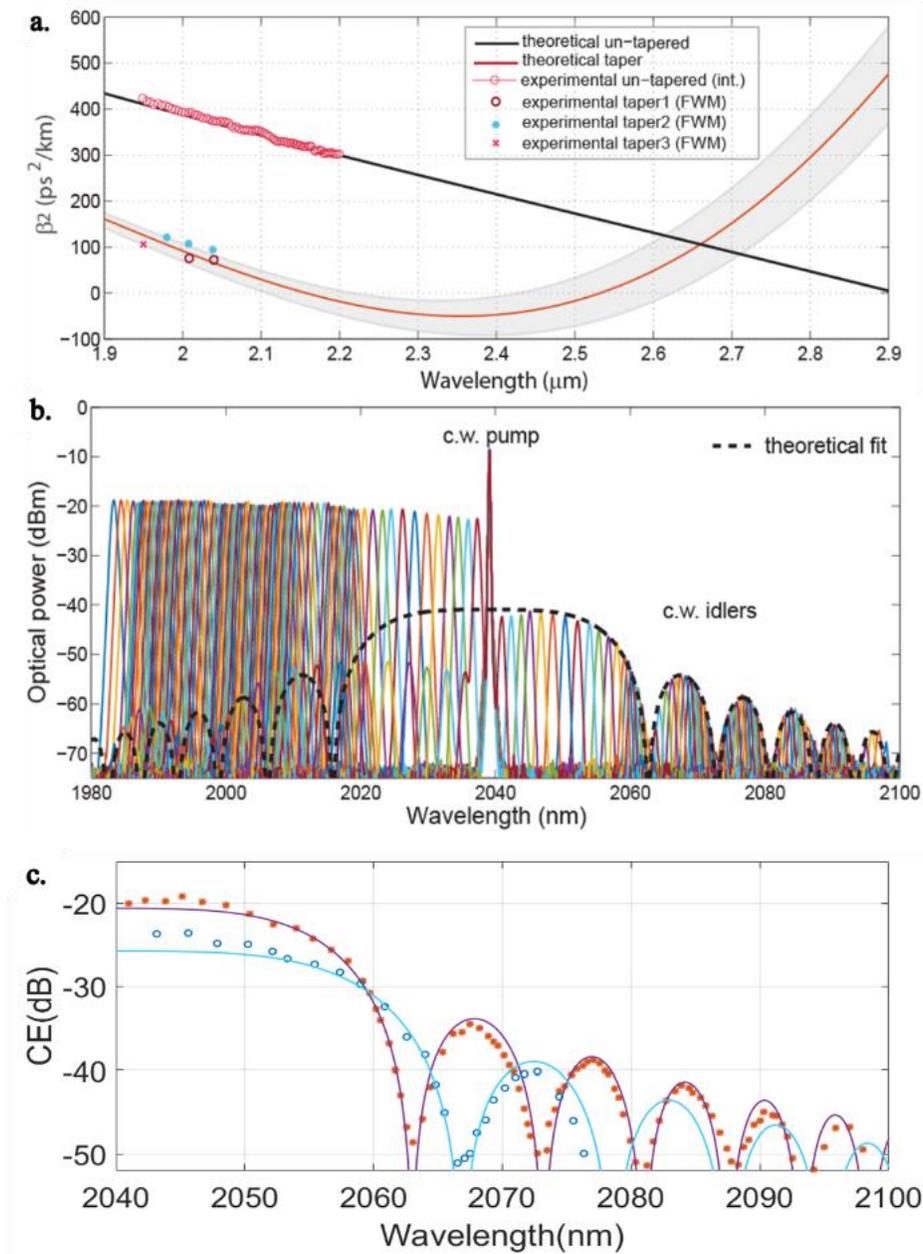

**Figure 3 | Engineered GeAsSe microstructured taper characteristics. a**. Dispersion of the un-tapered and tapered fiber obtained from simulations (full lines). Measured dispersion on the fabricated fibers obtained from interferometric measurements (int.) and four-wave mixing data (FWM) are also plotted. **b.** Example of FWM spectra at the output of the tapered fiber for a c.w. pump positioned at 2040 nm with 13 mW power. Clear idler generation with phase matching features are observed. The data are in excellent agreement with the theoretical fitting. **c.** fitting of the 78 cm long tapered fiber at 2040 nm pump; rotating the input lensed fiber revealed its birefringence. Here, the discrete points represent the experimental values while continuous lines come from the theoretical fittings. The plot showed here are the maximum and minimum conversion bandwidths found changing the input laser polarization.

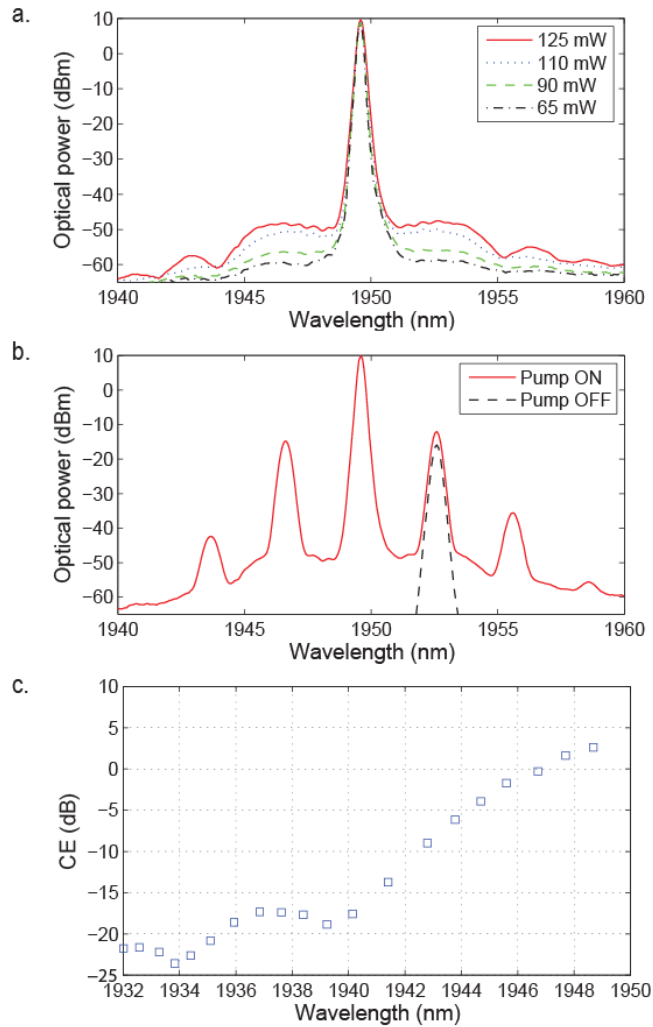

**Figure 4 | Experimental characterization of a 1m long taper for 1950 nm pump. a,** Output spectra for 4 coupled pump powers with no signal showing amplification of the ASE from the pump laser. **b.** Output spectrum for 125 mW pump and 1mW signal coupled power. CE above transparency, signal amplification and cascaded FWM is observed. **c**. Conversion efficiency as a function of idler wavelength for 125 mW pump power at 1950 nm.

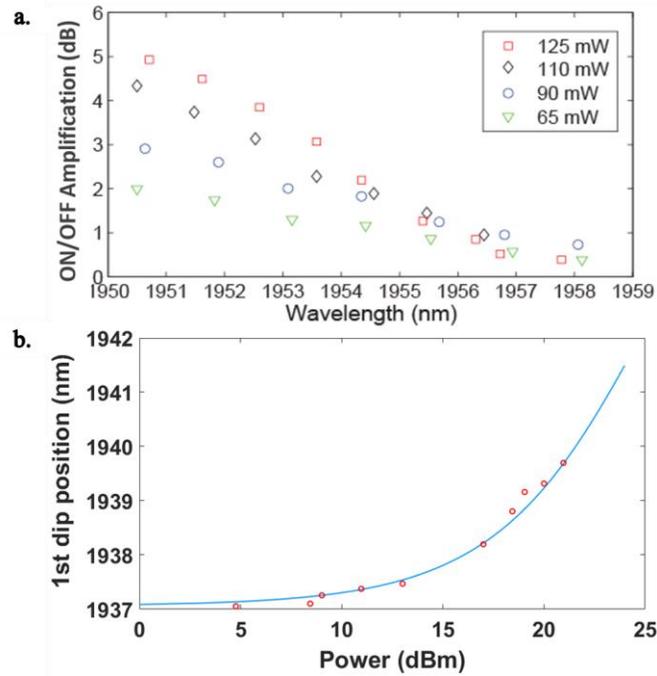

**Figure 5 | Experimental characterization of a 1m long taper for 1950 nm pump. a**. ON/OFF signal amplification for 4 coupled pump powers. Amplification is measured over close to 10 nm of bandwidth. A slight reduction in bandwidth happens with increasing pump power due to additional nonlinear phase mismatch. **b**. The experimentally measured and simulated positions of the CE's 1$^{st}$ dip as a function of coupled pump power. A perfect matching between experimental and theoretical value can be seen, which implies fiber structure was maintained during the high power test.

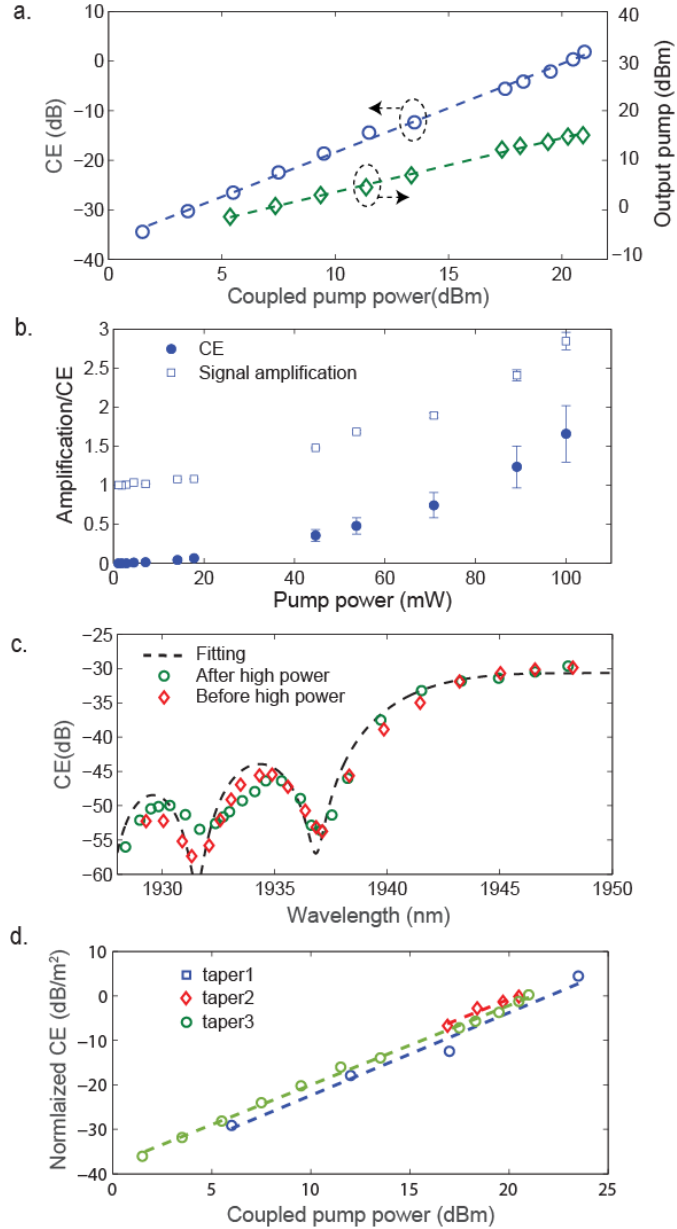

**Figure 6 |Mid-infrared parametric amplifier performance and repeatability. a.** Conversion efficiency as a function of coupled pump power for a 1950 nm pump and a 1953 nm signal. A slope of 2 is retrieved with no visible onset of saturation. The output pump power as a function of coupled pump power is also plotted, showing linear relationship. **b.** Amplification and CE as a function of pump power. **c**. Measured CE as a function of wavelength for a 3 mW pump before and after high power (125 mW) testing, and fitting curve. **d.** CE normalized to the fiber length squared as a function of coupled pump power for three different tapers obtained from the same fiber. Similar performances are measured, confirming the repeatability of the experiment and of the taper features. Note that taper3 was tested at 1980nm instead of 1950nm.